\documentstyle[epsfig,twocolumn,prl,aps]{revtex}

\newcommand{\be}{\begin{equation}}
\newcommand{\ee}{\end{equation}}
\newcommand{\bea}{\begin{eqnarray}}
\newcommand{\eea}{\end{eqnarray}}
\renewcommand{\d}{{\rm d}}
\newcommand{\lton}{\mathrel{\lower.9ex
                  \hbox{$\stackrel{\displaystyle <}{\sim}$}}}

\begin{document}                                                
%\draft
\title{How Protons Shatter Colored Glass}
\author{Adrian Dumitru$^a$ and Larry McLerran$^b$}
%\bigskip
\address{
a) Department of Physics, Columbia University, New York, New York 10027\\
{\small email: dumitru@nt3.phys.columbia.edu}\\
b) Department of Physics, Brookhaven National Laboratory,
Upton, New York 11973-5000\\
{\small email: mclerran@bnl.gov}\\
}
\date{\today}
\maketitle
\begin{abstract} 
%************
We consider the implications of the Color Glass Condensate for the central
region of $p+A$ collisions. We compute the $k_\perp$ distribution of radiated
gluons and their rapidity distribution $\d N/\d y$ analytically, both in the
perturbative regime and in the region between the two saturation momenta.
We find an analytic expression for the number of produced gluons which is
valid when the saturation momentum of the proton is much less than that of the 
nucleus. We discuss the scaling of the produced multiplicity with $A$. We show
that the slope of the rapidity density $\d N/\d y$ provides an experimental
measure for the renormalization-group evolution of the color charge density of
the Color Glass Condensate (CGC). We also argue that these results are easily
generalized to collisions of nuclei of different $A$ at central rapidity,
or with the same $A$ but at a rapidity far from the central region.
%************
\end{abstract}
%\begin{narrowtext}

\section{Introduction}

The color field of a strongly Lorentz boosted hadron can be
described as a classical color field~\cite{McLerran:1994ni}, so long
as one has a high enough density of gluons such that the field modes 
have very large occupation numbers. The typical transverse momentum scale
for which the field modes have large occupation number will be called 
$Q_s$, the saturation momentum.
Seen with a resolution scale of $Q_s$, the collision
of two hadrons (say pions, protons, or nuclei) at very high energy can be
viewed as two (highly Lorentz contracted)
sources of color charge propagating along the light-cone.
Renormalization group
evolution in rapidity leads to a longitudinal extension of the source.
That is, the charge distribution is spread out on a scale given by the
characteristic longitudinal momentum of the ``hard'' particles which
generate the source of color charge entering the Yang-Mills equation for the
``soft'' modes. 

The field in front of and behind each ``sheet'' of charge is a pure
gauge~\cite{McLerran:1994ni}.  The color electric and magnetic
fields associated with these pure gauge vector potentials vanish, except
in the sheet where the vector potential is discontinuous (on a scale larger
than the longitudinal spread of the color charge source).
When the two sheets collide, corresponding to the tip of the
light-cone, the two charge sheets interact.  This produces radiation
in the forward light cone.

The point of our paper is to compute this radiation for collisions of
particles with different saturation momentum scales.  This problem turns out 
to be more tractable than that of collisions of two particles with equal
saturation scales. For example, to compute the production of particles
in the central region of equal $A$ nuclear collisions, one must perform 
intensive numerical computations~\cite{KV}.  If one collides
protons with nuclei at very high energies and studies the central region of
particle production, there are two scales, the saturation momentum of the
proton and that of the nucleus.
In the limit where $\Lambda_{QCD} \ll Q_s^{\rm proton}\ll
Q_s^A$, we shall see that the problem simplifies, and one can obtain 
analytic results for quantities such as the total multiplicity density of
gluons at zero rapidity.

The saturation momentum squared is proportional to the total number
of gluons in the hadron wavefunction
at rapidities larger than that at which we compute the production of particles.
One could introduce an asymmetry in the saturation scales by considering
equal $A$ nuclear collisions far from the central rapidity region.  
Alternatively, one could consider collisions of different nuclei, or various
combinations of the above.  In this sense, the proton in the $p+A$
scattering case
which we consider should be thought of as a generic acronym for asymmetric
nuclear collisions in either baryon number or rapidity.
  
We shall specifically consider the situation where the source
propagating along the $x^+=(t+z)/\sqrt{2}$ axis is much weaker than that
propagating along the $x^-=(t-z)/\sqrt{2}$ axis. In such a case, the
saturation momentum scale $Q_s^{(1)}$ on which source one can be viewed as
a classical field is smaller than the corresponding scale for source
two, $Q_s^{(2)}$. This fact has a very interesting consequence.
Namely, we expect three distinct regions in transverse momentum.
At large transverse momentum, $k_\perp> Q_s^{(2)}$, both fields are
weak. Thus, perturbation theory should be a valid approximation in this
regime~\cite{Kovner:1995ts,Gyulassy:1997vt,Kovchegov:1997ke}.
On the other hand, for $Q_s^{(2)}>k_\perp> Q_s^{(1)}$, the field one
is weak, and can be treated perturbatively; but field two is ``saturated'',
that is, the field strength $F^{\mu\nu}={\cal O}(1/g)$ has attained maximum
strength~\cite{McLerran:1994ni,Mueller:1999wm},
and is in the non-linear regime.
In that regime of transverse momentum, field two can not be treated as a small
perturbation, even if $Q_s^{(2)}\gg\Lambda_{\rm QCD}$ and the coupling
$\alpha_s(Q_s^{(2)})\ll1$ is weak. Those non-linearities modify the
transverse momentum distribution of radiation produced due to the
interaction. Our goal here is to compute the distribution in the intermediate
regime $Q_s^{(2)}>k_\perp> Q_s^{(1)}$.
Finally, at an even smaller transverse momentum
$<Q_s^{(1)}$, both fields are strong. In this region we expect a flat
$k_\perp$ distribution, up to logarithms of $k_\perp^2$.
However, we can presently not compute the distribution in that region
analytically, but it has been obtained numerically~\cite{KV}.

\begin{figure}[htp]
\centerline{\hbox{\epsfig{figure=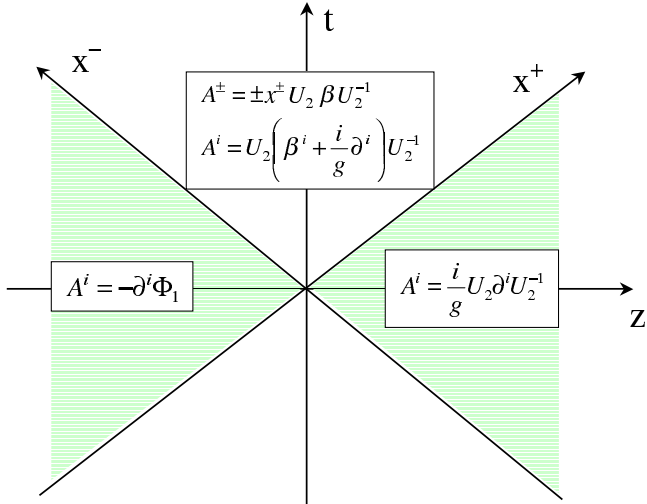,width=8.5cm}}}
\caption{The solutions of the Yang-Mills equations in the various
parts of the light-cone. The charge distributions propagate along the
$x^-$, $x^+$ axes. In the space-like regions behind the
charge distributions the fields are just
gauge transformations of vacuum fields, rotated by the respective
charge densities of the sources. In the forward light-cone, the field
at time $\rightarrow\infty$ is given by gauge rotated plane wave
solutions $\beta$ and $\beta^i$.}
\label{figLC}
\end{figure}
The solution of the nonabelian Yang-Mills equations is illustrated in
Fig.~\ref{figLC}. The two color-charge distributions propagate along the
$x^-$, $x^+$ axes. The fields $A^i$ in the space-like regions behind them
are just gauge rotated vacuum fields, and $A^\pm=0$ there.

For an abelian gauge group (electrodynamics) the field in the forward
light-cone is just the sum of the two pure-gauge fields behind the
propagating charge distributions. It is also just a gauge rotated vacuum
field, and so no radiation occurs (if recoil is neglected).
For a non-abelian gauge group
(chromodynamics) the sum of two pure gauges is not a pure gauge, and
radiation occurs at the classical level, even when recoil is neglected.
At asymptotic times, the
field in the forward light-cone must be given by gauge rotated plane waves.
In a leading order perturbative
computation~\cite{Kovner:1995ts,Gyulassy:1997vt,Kovchegov:1997ke} those
gauge rotations can be expanded to first order in the gauge potentials.
However, in order to reach into
the non-linear ``saturation regime'' we must account for the interaction of
the radiation field with the fields of the color-charge distributions on the
light-cone to all orders. A numerical approach to this problem has been
used in ref.~\cite{KV} for collisions of equal-size nuclei, and at
midrapidity. For current-nucleus interactions (Deep Inelastic Scattering
off large nuclei) the distribution function of produced gluons in the
fragmentation region has been obtained analytically
(via a diagrammatic approach) in~\cite{Kovchegov:1998bi}, where the
authors also discuss the generalization of their result to $p+A$ collisions.
In the central rapidity region (the region $z\ll t$ in Fig.~\ref{figLC})
one should also account for the
renormalization group (RG) evolution of the CGC color charge density per
unit transverse area~\cite{Iancu:2001ad,Jalilian-Marian:1997xn}.
The purpose of this paper is to derive analytically an explicit expression
for the transverse momentum and rapidity distribution of produced gluons
in $A_1+A_2$ collisions at high energy, valid at all rapidities
where the CGC color charge density including RG evolution is much larger
for the source $A_2$ than for $A_1$.
Our explicit result for $\d N/\d k_\perp^2\d y$
shows that the transverse momentum distribution is modified from a
$\sim1/k_\perp^4$ behavior in the perturbative regime (high $k_\perp$)
to $\sim1/k_\perp^2$ in the region where $k_\perp$ is between the
saturation scales for the two sources. Furthermore, we show that
the slope of the multiplicity per unit of rapidity, $\d N/\d y$, provides
an experimental measure for the RG evolution of the CGC color charge density.

This article is organized as follows. In section~\ref{calculation} we
derive the transverse momentum and rapidity distribution of the radiated
gluons to all orders in the field of the large nucleus. We do this by
solving the Yang-Mills equations with the appropriate boundary conditions.
This section is
somewhat technical and can be skipped by readers interested only in the
results relevant for phenomenology.
In section~\ref{discussion} we discuss the most important features of
the radiation spectrum in the perturbative regime (high transverse momentum)
and in the ``saturation regime'', including the $A$-scaling and the
evolution in rapidity. We outline possible ways of measuring experimentally
the RG evolution of the color charge density of the CGC. We summarize
in section~\ref{summary}.

\section{The Distribution of Produced Gluons} \label{calculation}
In this section, we shall first solve the Yang-Mills equations in
coordinate space. We assume that in the forward light-cone
the vector potential depends only on the transverse coordinate $x_\perp$ and
on proper time, $\tau=\sqrt{t^2-z^2}\equiv
\sqrt{2x^+x^-}$ (which is invariant under longitudinal Lorentz boosts),
but not on rapidity $y=\log(x^+/x^-)/2$. When performing the path integral over
the ``hard'' source for the classical color field in eq.~(\ref{Gauss}) we shall
explicitly consider the dependence on rapidity.

In the space-like regions the transverse fields are pure
gauges~\cite{McLerran:1994ni},
\bea
\alpha_m^i &=& -\frac{1}{ig} U_m(x_\perp) \partial^i U^\dagger_m(x_\perp)
                 \nonumber\\ &=& -\frac{1}{ig} e^{-ig\Phi_m(x_\perp)} \partial^i e^{ig\Phi_m(x_\perp)}
\quad (m=1,2)~. \label{defU}
\eea
The fields $\Phi_m$ satisfy
\bea
        - \nabla^2_\perp \Phi_m  = g \rho_m(x_\perp)~.
\eea
We take the distribution of the sources of the gluon color field for 
each nucleus  as a Gaussian according to the McLerran-Venugopalan model,
\bea
\int {\cal D}\rho_1\, {\cal D}\rho_2\; \exp\left(-F_1[\rho_1] - F_2[\rho_2]
   \right)~,
\eea
where
\bea
        F_i [\rho_i] = \int \d y \, \d^2x_\perp \;
{\rm tr}\, \rho_i^2(x_\perp,y)/\mu^2_i(y)~.
\eea
The quantity $\mu^2(y)$ is the color charge squared per
unit rapidity and per unit transverse area scaled by $N_c^2 -1$.
It can be related to the gluon distribution function with
known coefficients, as shown in Ref.~\cite{Gyulassy:1997vt}.
It will turn out that the radiation distribution depends only on integrals over
$\mu^2(y)$, i.e.\ the {\em total} color charge squared from
rapidities greater than that at which we are interested in computation.

The field $\Phi_1$ will be assumed to be weak such that the exponentials
in eq.~(\ref{defU}) can be expanded to leading order,
\be
\alpha_1^i = -\partial^i \Phi_1 + {\cal O}(\Phi_1^2)~.
\ee
In the forward light-cone we write the transverse and $\pm$ components of
the gauge field as
\bea
A^i(\tau,x_\perp) &=& \alpha_3^i(\tau,x_\perp)~,\\
A^\pm(\tau,x_\perp) &=& \pm x^\pm \alpha(\tau,x_\perp)~,
\eea
corresponding to the gauge condition
\be
x^+A^- + x^-A^+ =0~.
\ee
Thus, our ansatz for the gauge fields is
\bea
A^i(\tau,x_\perp) &=& \alpha_3^i(\tau,x_\perp)\Theta(x^-)\Theta(x^+)\nonumber\\
                     &+&\alpha_1^i(x_\perp)\Theta(x^-)\Theta(-x^+) \nonumber\\
                     &+&\alpha_2^i(x_\perp)\Theta(-x^-)\Theta(x^+) ~,\\
A^\pm(\tau,x_\perp) &=& \pm x^\pm \alpha(\tau,x_\perp)\Theta(x^-)\Theta(x^+)~.
\eea
Next, we determine the boundary conditions for $x^-,x^+\rightarrow 0$.
In that limit, 
\be
\left[ D_+, F^{+ i}\right] + \left[ D_-, F^{- i}\right] =2
\delta(x^-) \delta(x^+) \left( \alpha_3^i - \alpha_1^i
                               -\alpha_2^i\right)~.
\ee
(The contribution from $[D_j, F^{j i}]$ is not singular at $x^+=x^-=0$.)
For this term to vanish identically we must
satisfy the boundary condition
\be \label{bc_alphai}
\alpha_3^i(\tau=0,x_\perp) = \alpha_1^i(x_\perp) + \alpha_2^i(x_\perp)~,
\ee
as found before in~\cite{Kovner:1995ts,Gyulassy:1997vt}.

Using~(\ref{bc_alphai}), the equation
\be
\left[ D_\mu, F^{\mu +}\right] = J^+\equiv \delta(x^-)
\, g \rho_1(x_\perp)~,
\ee
where $\rho_1$ is the charge density per transverse area in nucleus 1, gives
for $x^+,x^-\rightarrow0$
\bea
& & \delta(x^-)\left\{ 2\alpha \Theta(x^+) +
 \partial_i\alpha_1^i - \Theta(x^+) \,
    ig \left[\alpha_1^i,\alpha_2^i\right]\right\}\nonumber\\
& =&  \delta(x^-) \, \partial_i\alpha_1^i~.
\eea
That requires the matching condition~\cite{Kovner:1995ts,Gyulassy:1997vt}
\be \label{bc_alpha}
\alpha(\tau=0,x_\perp) = \frac{ig}{2}\left[ \alpha_1^i(x_\perp),
                                             \alpha_2^i(x_\perp)\right]~.
\ee

We now determine the solution in the forward light-cone,
$x^+,x^- >0$. $[D_\mu,F^{\mu \nu}]=0$ becomes~\cite{Kovner:1995ts}
\bea
\frac{1}{\tau^3} \partial_\tau \tau^3 \partial_\tau \alpha -
   \left[D^i,\left[D_i,\alpha\right]\right] &=& 0~,\label{eqom1}\\
\frac{1}{\tau} \left[D_i,\partial_\tau\alpha_3^i\right] + ig\tau\left[
       \alpha,\partial_\tau\alpha\right] &=& 0~,\label{eqom2}\\
\frac{1}{\tau} \partial_\tau \tau \partial_\tau \alpha_3^i -
         ig\tau^2\left[\alpha,\left[D^i,\alpha\right]\right]-
         \left[D^j,F^{ji}\right] &=& 0~.\label{eqom3}
\eea
We assume that the field of the second nucleus is much stronger than the
radiation field, and so linearize the equations of motion in $\alpha$. (Note
that $\alpha\rightarrow 0$ if source one becomes arbitrarily weak, as no
radiation occurs in the single nucleus case.) That
amounts to dropping the second terms in eqs.~(\ref{eqom2},\ref{eqom3}).
We perform a gauge rotation
\bea
\alpha(\tau,x_\perp) &=& U_2(x_\perp) \beta(\tau,x_\perp)
                         U_2^\dagger(x_\perp)~, \\
\alpha_3^i(\tau,x_\perp) &=& U_2(x_\perp) \left( \beta^i(\tau,x_\perp) - 
                         \frac{1}{ig} \partial^i\right) U_2^\dagger(x_\perp)~,
\eea
with $U_2=\exp(-ig\Phi_2)$ as defined in~(\ref{defU}).
Then $[D^i,\cdot]$ becomes the ordinary derivative $\partial^i$ up to
corrections of order ${\cal O}(\beta^i)$ which do not show up in the
linearized equations of motion,
\bea
\frac{1}{\tau^3} \partial_\tau \tau^3 \partial_\tau \beta -
   \partial^i\partial_i\beta &=& 0~,\label{eqom1p}\\
\partial_\tau\partial_i\beta^i &=& 0~,\label{eqom2p}\\
\frac{1}{\tau} \partial_\tau \tau \partial_\tau \beta^i -
         \left(\partial^k\partial_k\delta^{ij}-\partial^i\partial^j
         \right)\beta^j &=& 0~.\label{eqom3p}
\eea
Gauge rotating the boundary conditions~(\ref{bc_alphai},\ref{bc_alpha}) gives
\bea
\beta^i(\tau=0,x_\perp) &=& U_2^\dagger(x_\perp) \alpha_1^i(x_\perp)
                          U_2(x_\perp)~,\\
\beta(\tau=0,x_\perp) &=& \nonumber\\
& & \hspace*{-1cm} \frac{ig}{2} U_2^\dagger(x_\perp) \left[
                           \alpha_1^i(x_\perp),\alpha_2^i(x_\perp)\right]
                           U_2(x_\perp)~.
\eea
From~(\ref{eqom2p}), $\partial_i\beta^i(\tau,x_\perp)$ is time independent.
We can thus write
\be
\beta^i(\tau,x_\perp) = \epsilon^{il}\partial^l \chi(\tau,x_\perp) +
                        \partial^i\Lambda(x_\perp)~,
\ee
where $\epsilon^{il}$ is the Levi-Civita tensor in two
dimensions. The first term contributes to the curl while the second contributes
to the divergence of $\beta^i$. This ansatz for $\beta^i$ makes~(\ref{eqom2p})
an identity. The equations of motion~(\ref{eqom1p}-\ref{eqom3p}) now read
\bea
\frac{1}{\tau^3} \partial_\tau \tau^3 \partial_\tau \beta -
   \partial^i\partial_i\beta &=& 0~,\label{eqom1pp}\\
\frac{1}{\tau} \partial_\tau \tau \partial_\tau \tilde\chi -
         \partial^i\partial_i \tilde\chi &=& 0~,\label{eqom3pp}
\eea
where $\tilde\chi = -\partial^j\partial_j \chi$. The boundary condition for
$\tilde\chi$ can be obtained by noting that $\tilde\chi=\epsilon^{ij}
\partial^j\beta^i$,
\be
\tilde\chi(\tau=0,x_\perp) = \epsilon^{ij} \partial^j U_2^\dagger(x_\perp)
                          \alpha_1^i(x_\perp) U_2(x_\perp)~.
\ee
The equations of motion~(\ref{eqom1pp},\ref{eqom3pp}) are solved by a
superposition of Bessel functions,
\bea
\beta(\tau,x_\perp) &=& \int\frac{\d^2k_\perp}{(2\pi)^2} e^{ik_\perp \cdot
                 x_\perp} b_1(k_\perp) \frac{1}{\omega\tau}J_1(\omega\tau)~,
               \label{sol_beta}\\
\tilde\chi(\tau,x_\perp) &=& \int\frac{\d^2k_\perp}{(2\pi)^2} e^{ik_\perp \cdot
                 x_\perp} b_2(k_\perp) J_0(\omega\tau)~. \label{sol_tchi}
\eea
The functions $b_1(k_\perp)$, $b_2(k_\perp)$ are determined in coordinate
space by the boundary conditions for the fields at $\tau=0$:
\bea
b_2(x_\perp) &=& \int\frac{\d^2k_\perp}{(2\pi)^2} e^{ik_\perp \cdot x_\perp} 
                 b_2(k_\perp) = \tilde\chi(\tau=0,x_\perp)~,\\
b_1(x_\perp) &=& \int\frac{\d^2k_\perp}{(2\pi)^2} e^{ik_\perp \cdot x_\perp} 
                 b_1(k_\perp) = 2\beta(\tau=0,x_\perp)~.
\eea
This follows from the expansion of $J_\nu(x)$ for small $x$:
$J_\nu(x) \simeq (x/2)^\nu/\nu!$.
Asymptotically, for $x\rightarrow\infty$, the Bessel functions are
$J_\nu(x) \simeq \sqrt{2/\pi x} \cos(x-\nu\pi/2-\pi/4)$.
Also, for time $\tau\rightarrow\infty$ we assume free fields,
$\omega=|\vec{k}_\perp|$.
Then, comparing the solutions~(\ref{sol_beta},\ref{sol_tchi}) for
$\tau\rightarrow\infty$ to plane waves~\cite{Kovner:1995ts},
\bea
\beta(\tau\rightarrow\infty,x_\perp) &=& \int\frac{\d^2k_\perp}{(2\pi)^2}
              \frac{1}{\sqrt{2\omega\tau^3}}\nonumber\\
          & & \times \left\{
           a_1(k_\perp)e^{ik_\perp \cdot x_\perp-i\omega\tau} +c.c.\right\}~,\\
\beta^i(\tau\rightarrow\infty,x_\perp) &=& \int\frac{\d^2k_\perp}{(2\pi)^2}
\frac{1}{\sqrt{2\omega\tau}}\frac{\epsilon^{il}k_\perp^l}{\omega}\nonumber\\
          & & \times  \left\{
          a_2e^{ik_\perp \cdot x_\perp-i\omega\tau} +c.c.\right\}~,
\eea
yields for purely imaginary $b_2(k_\perp)$ and real $b_1(k_\perp)$:
\bea
a_1(k_\perp) &=& \frac{1}{\sqrt{\pi}k_\perp} b_1(k_\perp)
                 e^{3\pi i/4} \nonumber\\
             &=& \frac{ig}{\sqrt{\pi}k_\perp}e^{3\pi i/4} \int\d^2x_\perp
                 e^{-ik_\perp \cdot x_\perp} \nonumber\\
           & &\times U_2^\dagger(x_\perp) \left[
                           \alpha_1^i(x_\perp),\alpha_2^i(x_\perp)\right]
                           U_2(x_\perp)~,\label{eq_a1}\\
a_2(k_\perp) &=& \frac{1}{\sqrt{\pi}k_\perp} b_2(k_\perp)
                 e^{i\pi/4} \nonumber\\
             &=& \frac{i}{\sqrt{\pi}k_\perp}e^{i\pi/4} \int\d^2x_\perp
                 e^{-ik_\perp \cdot x_\perp} \nonumber\\
             & &\times \epsilon^{ij} \partial^j U_2^\dagger(x_\perp)
                          \alpha_1^i(x_\perp) U_2(x_\perp)~.
\eea
To simplify $a_1$, recall from~(\ref{defU}) that
$\alpha_2^i=U_2(-1/ig)\partial^i U_2^\dagger$. Therefore,
\bea \label{tra1}
ig U_2^\dagger \left[\alpha_1^i,\alpha_2^i\right] U_2 &=&
 \partial^i U_2^\dagger \alpha_1^i U_2-
U_2^\dagger\left(\partial^i\alpha_1^i\right) U_2\nonumber\\
&=& \alpha_1^{a,i} \partial^i U_2^\dagger t^a U_2~.
\eea
Let us evaluate ${\rm tr} |a_2|^2$ first.
Squaring the amplitude and taking the trace yields
\bea
{\rm tr} |a_2|^2 &=& \frac{1}{{\pi}k_\perp^2} \int \d^2x_\perp\d^2z_\perp
e^{-ik_\perp \cdot (x_\perp-z_\perp)}      \nonumber\\
 & & \hspace{-1cm}\times \epsilon^{ij} \epsilon^{kl} \partial_x^j \partial_z^l 
   \,{\rm tr}\langle A^i(x_\perp,y) A^k(z_\perp,y)
     \rangle_{\Phi_1,\Phi_2}~. \label{tra2}
\eea
Here, $y$ denotes the rapidity.
The averaging is with respect to the gauge potentials $\Phi_1$ and $\Phi_2$,
assuming a Gaussian weight~\cite{McLerran:1994ni,Jalilian-Marian:1997xn}:
\bea \label{Gauss}
\langle O\rangle_\Phi &=& \int \!\! {\cal D} \Phi \, O(\Phi)\nonumber\\
&\times& \exp \left[
-\int
\d y'\int \d^2x_\perp \frac{{\rm tr} \left(\nabla^2_\perp
\Phi(x_\perp,y')\right)^2}
{g^2\mu^2(x_\perp,y')}\right]~.
\eea
When averaging over $\Phi_2$, the $y'$-integral extends from 
$-\infty$ (or some large negative rapidity beyond which the source vanishes)
to the rapidity of the produced gluons, $y$.
Vice versa, when averaging over $\Phi_1$ it goes
from $y$ to $+\infty$.

We now have to compute the correlation function
\be \label{averageA}
{\rm tr} \langle A^i(x_\perp,y) A^k(z_\perp,y) \rangle_{\Phi_1,\Phi_2}
\ee
with
\be
A^i(x_\perp,y) = U_2^\dagger(x_\perp,y)
 \left(\partial^i\Phi_1(x_\perp,y)\right)
 U_2(x_\perp,y)~.
\ee
The average over $\Phi_1$ in eqs.~(\ref{tra2},\ref{averageA})
 can be performed right away.
From~(\ref{Gauss}) we have~\cite{Jalilian-Marian:1997xn}
\bea
& & \partial_x^i \partial_z^k
\langle \Phi_1^a(x_\perp,y) \Phi_1^b(z_\perp,y) \rangle_{\Phi_1} 
\nonumber\\ 
&=& g^2 \delta^{ab} \int^{\infty}_y \!\! \d y' \,
     \mu_1^2(y') \,
     \partial_x^i \partial_z^k \gamma(u_\perp)
\nonumber\\ 
&=& g^2 \delta^{ab} \chi_1(y) \, \partial_x^i \partial_z^k 
\gamma(u_\perp)~, \label{2point}
\eea
with $u_\perp\equiv  x_\perp-z_\perp$. Also, we 
defined the total charge squared at rapidity $y$ induced by
the source from rapidities $[y,\infty]$ (not to be confused with the
auxilliary fields $\chi$, $\tilde\chi$ used above in intermediate steps
of the calculation),
\be
\chi_1(y) = \int^{\infty}_y \d y' \, \mu_1^2(y')~.
\ee

The tadpole-subtracted propagator is~\cite{Jalilian-Marian:1997xn}
\be \label{tad_prop}
\gamma(x_\perp)
= \frac{1}{8\pi} x_\perp^2 \log x_\perp^2\Lambda_{\rm QCD}^2~.
\ee
Also, in~(\ref{2point}) we assumed slow variation of $\mu_1$ over the
relevant transverse scales, and so neglect derivatives of it.

We are left with
\be \label{phi2_correlfct}
{\rm tr} \langle U^\dagger_2(x_\perp,y)
           t^a U_2(x_\perp,y)
U^\dagger_2(z_\perp,y)
           t^a U_2(z_\perp,y) 
     \rangle_{\Phi_2}~.
\ee
The most efficient way to evaluate this expression is to note that
\bea
& &\left( U^\dagger_2(x_\perp,y) t^a 
          U_2(x_\perp,y)\right)_{\alpha\beta}
 = \left( t^{a'}\right)_{\alpha\beta} U_{\rm adj}^{a'a}
(x_\perp) =\nonumber\\
& & \left( t^{a'}\right)_{\alpha\beta} \left( {\cal P}
\exp\left( ig\int^y_{-\infty}\!\!\d y' \,T^b
\Phi_2^b(x_\perp,y')\right)\right)^{a'a}~.
\eea
The path ordered exponential can be expanded as
\bea
1 &+& ig\int^y_{-\infty}\d y' T^b_{a'a}
\Phi_2^b(x_\perp,y') \nonumber\\
&+& (ig)^2 \int^y_{-\infty}\!\!\!\!\d y' \int^{y'}_{-\infty}\!\!\!\!\d y''
T^b_{a'd} T^c_{da} \Phi_2^b(x_\perp,y') \Phi_2^c(x_\perp,y'')\nonumber\\
&+& \cdots \label{pathOexp}
\eea
According to~(\ref{phi2_correlfct}) we have to multiply two such expressions,
one at $x_\perp$ and the other at $z_\perp$.
The zeroth order is of course trivial.
The contribution to ${\cal O}(g^2)$ arises from the product of the
two ${\cal O}(g)$ terms in~(\ref{pathOexp})
because tadpoles only enter via a subtraction of the
propagator, $\gamma(u_\perp)$, at
$u_\perp=0$~\cite{Jalilian-Marian:1997xn}.
We find
\bea
& &
  (ig)^2 \!\!\! \int^y_{-\infty}\!\!\!\! \d y' \!\!\!
  \int^{y}_{-\infty}\!\!\!\! \d \bar{y}'\,
  T^b_{a'a}T^{b'}_{a'a}\left\langle \Phi_2^b(x_\perp,y') 
  \Phi_2^{b'}(z_\perp,\bar{y}') 
\right\rangle_{\Phi_2} \nonumber\\
&=& {g^2} N_c \, \delta^{bb'} \!\!\! \int^y_{-\infty}\!\!\!\!\d y' 
\!\!\!\int^{y}_{-\infty}\!\!\!\! \d \bar{y}'\left\langle \Phi_2^b(x_\perp,y')
\Phi_2^b(z_\perp,\bar{y}') \right\rangle_{\Phi_2} \nonumber\\
&=& {g^4} N_c \, \delta^{bb'} \gamma(u_\perp)
 \int^y_{-\infty}\!\!\!\!\d y' \mu_2^2(y') 
     \nonumber\\
&=& {g^4} N_c \, \delta^{bb'} \gamma(u_\perp) \chi_2(y)~.
\eea
Analogously to the definition of $\chi_1(y)$ above, $\chi_2(y)$
denotes the total charge squared at rapidity $y$ induced by
the source from rapidities $[-\infty,y]$,
\be
\chi_2(y) = \int^y_{-\infty} \d y' \, \mu_2^2(y')~.
\ee

Next, we multiply two terms of ${\cal O}(g^2)$ from eq.~(\ref{pathOexp}).
Again, besides a subtraction at $u_\perp=0$ tadpole diagrams can be
disregarded, and so this is the only contribution to that order.
\bea
& &
  (ig)^4 \int^y_{-\infty}\!\!\!\!\d y' \int^{y'}_{-\infty}\!\!\!\!\d y''
\int^y_{-\infty}\!\!\!\!\d \bar{y}' \int^{\bar{y}'}_{-\infty}\!\!\!\!
\d \bar{y}''  \nonumber\\
&\times& T^b_{a'd} T^c_{da} T^{b'}_{a'd'} T^{c'}_{d'a} \nonumber\\
&\times& \left\langle
 \Phi_2^b(x_\perp,y') \Phi_2^c(x_\perp,y'')
 \Phi_2^{b'}(z_\perp,\bar{y}') 
 \Phi_2^{c'}(z_\perp,\bar{y}'')\right\rangle_{\Phi_2}~.
\eea
We can now contract $\Phi_2^b$ with
$\Phi_2^{b'}$ (and accordingly $\Phi_2^c$ with $\Phi_2^{c'}$); or we can
contract $\Phi_2^b$ with $\Phi_2^{c'}$
(and accordingly $\Phi_2^{c}$ with $\Phi_2^{b'}$). However, the latter
is zero because of the ordering in rapidity. Thus, we obtain
\bea
& & {g^8} N_c^2
\gamma^2(u_\perp)
\int^y_{-\infty}\!\!\!\!\d y' \int^{y'}_{-\infty}\!\!\!\!\d y''
\mu_{2}^2(y')\mu_{2}^2(y'') \nonumber\\
&=& {g^8} N_c^2 \gamma^2(u_\perp)
 \frac{1}{2!}\chi_2^2(y)~.
\eea
One can repeat the above steps to any order.
Resumming the series and summing over the one remaining adjoint color index
we find for eq.~(\ref{phi2_correlfct})
\be
\frac{N_c^2-1}{2}\exp\left\{g^4N_c\gamma\left(u_\perp\right)
\chi_2(y)\right\}~.
\ee
The correlation function~(\ref{averageA}) reads
\bea
& & \frac{N_c^2-1}{2}
g^2 \chi_1(y)  \left[\partial^i_x \partial^k_z
\gamma(u_\perp)\right]\nonumber\\
&\times& \exp\left\{g^4N_c \gamma\left(u_\perp\right)
\chi_2(y)\right\}~. \label{res_corfct}
\eea
From~(\ref{tra2}), $\epsilon^{ij} \epsilon^{kl}
\partial_x^j \partial_z^l$ acts on~(\ref{res_corfct}). 
The direct product with $\partial^i_x \partial^k_z\gamma(u_\perp)$ gives
zero, such that
effectively $\epsilon^{ij} \epsilon^{kl}
\partial_x^j \partial_z^l$ acts on the exponential only.

Using~(\ref{tra1}) in~(\ref{eq_a1}) one derives a very similar result for
${\rm tr}|a_1|^2$, with the replacement $\epsilon^{ij} \epsilon^{kl}
\partial_x^j \partial_z^l\rightarrow
\delta^{ij} \delta^{kl}\partial_x^j \partial_z^l$, and where again these
derivatives act on the exponential only.
In total we obtain
\bea
{\rm tr}& &\left(|a_1|^2+|a_2|^2\right) =
  \frac{N_c^2-1}{2\pi k_\perp^2} 
 \int\d^2x_\perp \d^2z_\perp e^{-ik_\perp \cdot u_\perp} \nonumber\\
& & \times g^2 \chi_1(y) \left[\partial_x^i \partial_z^k
\gamma(u_\perp)\right] \left(\epsilon^{ij} \epsilon^{kl}+
  \delta^{ij} \delta^{kl}\right) \nonumber\\
& &\times\partial_x^j \partial_z^l \exp\left\{g^4N_c\gamma(u_\perp)\chi_2(y)
\right\}~. \label{tot_trace}
\eea
[Aside: At this point, it is easy to verify that
the perturbative result obtained previously
in~\cite{Kovner:1995ts,Gyulassy:1997vt,Kovchegov:1997ke,Guo:1999pe}
is recovered when expanding the exponential to first order.
Using
\bea
& &\left(\epsilon^{ij} \epsilon^{kl}+\delta^{ij} \delta^{kl}\right)
 \left[\partial_x^i \partial_z^k \gamma(u_\perp)\right]
 \left[\partial_x^j \partial_z^l \gamma(u_\perp)\right] \nonumber\\
&=& 
\left\{
\int\frac{\d^2p_\perp}{(2\pi)^2} \frac{e^{ip_\perp\cdot u_\perp}}{p_\perp^2}
\right\}^2~,
\eea
the integral over $\d^2 u_\perp$ gives
$(2\pi)^2\delta(p_\perp+p'_\perp-k_\perp)$, while the integral over
$\d^2b_\perp\equiv\d^2(x_\perp+z_\perp)/2$ gives the transverse area $S_\perp$.
Thus,
\bea
\frac{\d N}{\d^2k_\perp\d y} &=& \frac{2}{(2\pi)^2} 
           {\rm tr}\left(|a_1|^2+|a_2|^2\right) \nonumber\\
& & \hspace{-1.5cm}
= S_\perp
  \frac{2 g^6 N_c(N_c^2-1)}{(2\pi)^3 k_\perp^2}
\chi_1 \chi_2 \int \frac{\d^2p_\perp}{(2\pi)^2} 
\frac{1}{p_\perp^2 (p_\perp-k_\perp)^2}~.
\eea
This result coincides with those of~\cite{Gyulassy:1997vt},
eq.~(36); \cite{Kovchegov:1997ke}, eq.~(40).
The remaining integral has to be regularized by introducing a
finite color neutralization correlation scale $\Lambda^2$, and
can then be written as
$k_\perp^{-2}\log(k_\perp^2/\Lambda^2)$~\cite{Gyulassy:1997vt}.
For the perturbative regime, that cutoff scale can be chosen as
$\Lambda^2=g^4N_c\chi_2/8\pi$.]

To simplify eq.~(\ref{tot_trace}) further, note that
\be
\left(\epsilon^{ij} \epsilon^{kl}+
  \delta^{ij} \delta^{kl}\right) 
\left[\partial^i_x \partial^k_z A\right]
\left[\partial^j_x \partial^l_z B\right]\nonumber\\
= \left[ \partial^2_x A\right] \left[\partial^2_x B\right]~,
\ee
as can be verified most easily in 2-d transverse Fourier space:
$(p_\perp\times q_\perp)^2 + (p_\perp\cdot q_\perp)^2 = 
p_\perp^2q_\perp^2(\cos^2(\phi)+\sin^2(\phi))=
p_\perp^2q_\perp^2$.
From the definition of $\gamma(u_\perp)$,
see eq.~(\ref{tad_prop}), we have $\partial^2\gamma(u_\perp)=(2+\log(u_\perp^2
\Lambda_{\rm QCD}^2))/2\pi$, and thus
\bea
\frac{\d N}{\d^2k_\perp\d y} &=&
  2\frac{N_c^2-1}{(2\pi)^4 k_\perp^2} 
 \int\d^2b_\perp \d^2u_\perp e^{-ik_\perp \cdot 
         u_\perp} g^2 \chi_1(y) \nonumber\\
& & \hspace*{-1.5cm} \times \left(2+\log(u_\perp^2\Lambda_{\rm QCD}^2)\right)
\partial^2 \exp\left\{g^4N_c\gamma(u_\perp)\chi_2(y)
\right\}~.\nonumber\\
& &
\eea
We can now integrate by parts. We neglect derivatives of the distribution
function of the small nucleus, i.e.\ of $\chi_1$, and of
the logarithm from the propagator. Then,
\bea
\frac{\d N}{\d^2k_\perp\d y} &=&
  2g^2\chi_1(y) \frac{N_c^2-1}{(2\pi)^4} 
 \int\d^2b_\perp \d^2u_\perp \, e^{-ik_\perp \cdot u_\perp} \nonumber\\
& & \hspace*{-1.5cm} \times \left(-2-\log(u_\perp^2\Lambda_{\rm QCD}^2)\right)
 \exp\left\{g^4N_c\gamma(u_\perp)\chi_2(y)
\right\}~, \label{main_result}
\eea
which is just the Fourier transform of 
\bea
\frac{\d N}{\d^2u_\perp\d y} &=&
  2g^2\chi_1(y) \frac{N_c^2-1}{(2\pi)^4} 
  \left(-2-\log(u_\perp^2\Lambda_{\rm QCD}^2)\right)\nonumber\\
& & \hspace*{-1.5cm} \times \int\d^2b_\perp 
\exp\left\{g^4N_c u^2_\perp\log(u_\perp^2\Lambda_{\rm QCD}^2)
  \chi_2(y)/8\pi\right\}~.\label{main_result_coord}
\eea
This is our main result. Eq.~(\ref{main_result}) gives the $k_\perp$
and $y$-distribution of produced gluons in the McLerran-Venugopalan model,
including the renormalization-group evolution of
$\chi$~\cite{Iancu:2001ad,Jalilian-Marian:1997xn,Balitsky:1996ub}.

In ref.~\cite{Kovchegov:1998bi} it was assumed
that nucleus 2 represents a uniform distribution of charge extending along
the rapidity axis from $y_0$ to $y_1$, such that $\chi_2(y_0)=0$. In other
words, neglect QCD evolution of $\chi_2$ and set $\chi_2(y)=
\mu_2^2 (y-y_0)$, with $\mu_2=const$. Then,
integrating over rapidity from $y_0$ to $y_1$ one obtains with logarithmic
accuracy at $x^2_\perp\ll1/\Lambda_{\rm QCD}^2$
\bea
\frac{\d N}{\d^2x_\perp} &\propto&
  \frac{N_c^2-1}{N_c g^2 x^2_\perp} \nonumber\\
& & \hspace*{-1.5cm} \times \int\d^2b_\perp 
\left[1-\exp\left\{ -g^4 N_c x^2_\perp\mu^2_2(y_1-y_0)/8\pi
\right\}\right]~.
\eea
With $\mu_2^2=2\pi^2\rho_{rel}xG(x)/g^2(N_c^2-1)$ one reproduces
the result of~\cite{Kovchegov:1998bi}. Here, $\rho_{rel}$ denotes the
(Lorentz-boosted) density of nucleons in nucleus 2:
\be
\rho_{rel} = \frac{\gamma A}{\pi R_A^2 (y_1-y_0)}~.
\ee

\section{Discussion} \label{discussion}
In this section we discuss the transverse momentum distribution of gluons
in various regimes, and the scaling of the multiplicity per unit of rapidity
with $A_1$ and $A_2$. When referring to the scaling with the mass numbers of
the two colliding nuclei, we shall specifically assume that at fixed
rapidity the color charge densities $\chi_i(y)$ are proportional to
$A_i^{1/3}$~\cite{Gyulassy:1997vt}. 

We can understand some general properties of
eqs.~(\ref{main_result},\ref{main_result_coord})
even without solving for the RG evolution of
$\chi(y)$. In the region where $x_\perp^2\Lambda^2_{\rm QCD}\ll
x_\perp^2g^4N_c\chi_2(y)/8\pi\ll1$, 
or alternatively $\Lambda^2_{\rm QCD}/k_\perp^2\ll
g^4N_c\chi_2(y)/8\pi k_\perp^2\ll1$, 
one can expand the exponential to first order
(the zero'th order term does not contribute to $k_\perp>0$).
Using
\bea
-2-\log(x_\perp^2\Lambda_{\rm QCD}^2) &=& \int\frac{\d^2p_\perp}{2\pi}
\frac{e^{i p_\perp\cdot x_\perp}}{p_\perp^2}~, \label{deriv_gamma}\\
\gamma(x_\perp) &=& \int\frac{\d^2q_\perp}{(2\pi)^2}
\frac{e^{i q_\perp\cdot x_\perp}}{q_\perp^4}~,
\eea
the integral over $\d^2u_\perp$ in eq.~(\ref{main_result}) just gives
$(2\pi)^2\delta(p_\perp+q_\perp-k_\perp)$, and we obtain
\bea
\frac{\d N}{\d^2b_\perp \d^2k_\perp\d y} &=& \nonumber\\
& &\hspace{-1.5cm}
 \frac{2g^6N_c(N_c^2-1)}{(2\pi)^4} \frac{\chi_1(y)\chi_2(y)}{k_\perp^4}
\log\frac{k_\perp^2}{g^4N_c\chi_2/8\pi}~. \label{pert_kt4}
\eea
Thus, one recovers the standard perturbative $\sim1/k_\perp^4$ behavior at 
very high $k_\perp$, with a logarithmic correction analogous to
DGLAP evolution~\cite{Gyulassy:1997vt,Guo:1999pe}.
Note that $\chi_1$, $\chi_2$ scale as $A_1^{1/3}$ and
$A_2^{1/3}$~\cite{Gyulassy:1997vt}, respectively,
while the integral over $\d^2b_\perp$ gives
a factor of $\pi R^2_2\propto A_2^{2/3}$. Therefore, in this kinematic
region $\d N/\d^2k_\perp\d y$ scales like $A_1^{1/3} A_2$, up to
logarithmic corrections. This holds also
for the integrated distribution $\d N(k_\perp>p_0)/\d y$ above some
fixed $A_2$-independent scale $p_0$.
On the other hand, when integrating over $k_\perp^2$ from
$g^4N_c\chi_2/8\pi$ to infinity, the contribution from large $k_\perp$ to
the rapidity density is
\be \label{dNdy_pert}
\frac{\d N}{\d^2b_\perp \d y} = \frac{g^2 (N_c^2-1)}{\pi^2}
 \chi_1(y)~.
\ee
Again, the integral over $\d^2b_\perp$ gives a factor $\pi R_2^2\propto
A_2^{2/3}$, and so $\d N/\d y$ scales like
$A_1^{1/3} A_2^{2/3}$; in this regard, see also the discussion
in~\cite{blaizot}.
The transverse energy can be obtained
from $\d E_\perp=k_\perp\d N$, using the number distribution~(\ref{pert_kt4}):
\be \label{Etpert}
\frac{\d E_\perp}{\d^2b_\perp\d y} = g^4
\sqrt{\frac{2N_c}{\pi^5}}(N_c^2-1)\chi_1(y)\sqrt{\chi_2(y)}~.
\ee
Finally, from~(\ref{pert_kt4}) and~(\ref{dNdy_pert})
the average transverse momentum in the perturbative regime is
\be
\langle k_\perp\rangle = 4 \, Q_s^{(2)}~,
\ee
where $Q_s^{(2)}(y)\equiv\sqrt{g^4N_c\chi_2(y)/8\pi}$.

\begin{figure}[htp]
\centerline{\hbox{\epsfig{figure=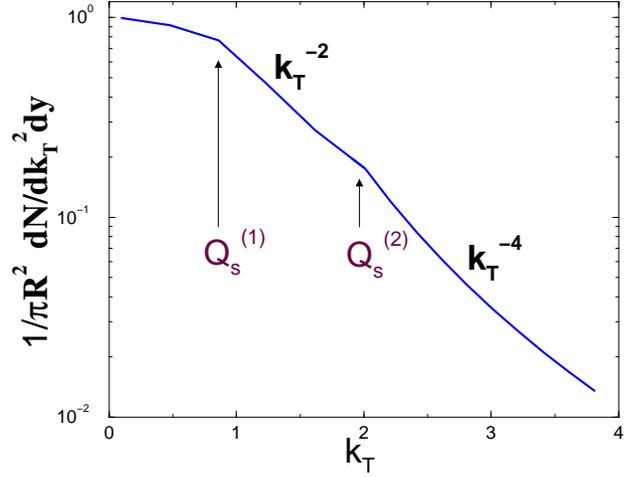,width=8cm}}}
\caption{Schematic $k_\perp$ distribution for particles produced in
high-energy $p+A$ collisions (or, more generally, for particles produced
in $A_1+A_2$
collisions at rapidity $y$ such that $\chi_1(y)\ll\chi_2(y)$).
In the perturbative regime, $\d N/\d k_\perp^2\d y\sim1/k_\perp^4$.
Inbetween the saturation scales for the two sources,
$\d N/\d k_\perp^2\d y\sim1/k_\perp^2$.}
\label{figkt}
\end{figure}

Within
the ``saturation regime'', i.e.\ when
$k_\perp^2< g^4N_c\chi_2(y)/8\pi$ but
$k_\perp^2>g^4N_c\chi_1(y)/8\pi$, the upper limit on the integral over
$u_\perp$ in eq.~(\ref{main_result})
is effectively given by $u_\perp^2<8\pi/g^4N_c\chi_2(y)$.
That is because the exponential suppresses contributions from larger
$u_\perp^2$. For the derivative of the propagator we may again use
eq.~(\ref{deriv_gamma}).
Then we find
\bea
\frac{\d N}{\d^2b_\perp \d^2 k_\perp \d y}
   &\simeq& 2g^2\chi_1(y)\frac{N_c^2-1}{(2\pi)^4}\nonumber\\
& &\hspace*{-1cm} \times \int\frac{\d^2 p_\perp}{2\pi p_\perp^2} 
 \int\limits^{8\pi/g^4N_c\chi_2}
\d^2 u_\perp e^{-i(k_\perp-p_\perp)\cdot u_\perp} \nonumber\\
&\simeq& g^2\chi_1(y)\frac{N_c^2-1}{(2\pi)^3}
 \int\frac{\d^2 p_\perp}{2\pi} \frac{1}{p_\perp^2(k_\perp-p_\perp)^2}
 \nonumber\\
&\simeq& g^2\chi_1(y)\frac{N_c^2-1}{(2\pi)^3}
 \frac{1}{k_\perp^2}\log\frac{k_\perp^2}{g^4N_c\chi_1(y)/8\pi}~.
\nonumber\\ \label{sat_kt2}
\eea
This form $\propto \chi_1/k_\perp^2$ is to be compared with that from
eq.~(\ref{pert_kt4}), $\propto \chi_1\chi_2/k_\perp^4$, valid at high
$k_\perp$. 
A schematic distribution\footnote{We mention again that we are in fact not
able to compute the distribution below $Q_s^{(1)}$. It has to be obtained
numerically using the methods of~\cite{KV}. In Fig.~\ref{figkt} we only
express the qualitative expectation that the distribution eventually flattens
out at very small $k_\perp$~\cite{KV,Kovner:1995ts,Mueller:1999wm}.}
in transverse momentum is shown in Fig.~\ref{figkt},
where $Q_s^{(i)}$ stands for $\sqrt{g^4N_c\chi_i(y)/8\pi}$.

Using the DGLAP equation for the transverse evolution, we can also
express the logarithm times the $\chi_1$ in eq.~(\ref{sat_kt2})
in terms of the unintegrated gluon distribution
function~\cite{Gyulassy:1997vt,Kovchegov:1998bi}. We write
\be
\chi_1(y) = \frac{1}{\pi R_1^2} \frac{N_c}{N_c^2-1} N^g_1(x,p_\perp^2)
\ee
and move the gluon number
\bea
\frac{\alpha_s N_c}{\pi} N^g_1(x,p_\perp^2) &=&
\frac{\alpha_s N_c}{\pi} \int_x^1\d x' G(x',p_\perp^2) \nonumber\\
&\approx& \frac{\d}{\d\log p_\perp^2} xG(x,p_\perp^2)
\eea
inside the integral over $\d^2p_\perp$ in eq.~(\ref{sat_kt2}).
This leads to
\bea
\frac{\d N}{\d^2b_\perp \d^2 k_\perp \d y} &\simeq&
\frac{1}{\pi R_1^2} \frac{1}{2\pi k_\perp^2} 
\int\limits^{k_\perp^2}_{g^4N_c\chi_1/8\pi} \!\!\! \d p_\perp^2
\frac{\d}{\d p_\perp^2} xG(x,p_\perp^2) \nonumber\\
& &\hspace*{-2cm} = \frac{1}{\pi R_1^2} \frac{1}{2\pi k_\perp^2} 
\left[ xG(x,k_\perp^2)-xG(x,g^4N_c\chi_1/8\pi)\right]~.
\eea
A quantitative computation of the radiation distribution requires to
determine numerically
the CGC density scales $\chi_1(y)$ and $\chi_2(y)$ from some
parametrization of the gluon and quark/antiquark distribution functions.
Also, the fragmentation of the radiated gluons into pions must be
taken into account. We postpone those issues to a future publication.

From~(\ref{sat_kt2}), the $k_\perp$-integrated multiplicity in the
{\em nonperturbative} regime
$g^4N_c\chi_1(y)/8\pi \lton k_\perp^2\lton g^4N_c\chi_2(y)/8\pi$ is
\be \label{dNdy_sat}
\frac{\d N}{\d^2b_\perp \d y} =
\frac{1}{4} g^2\chi_1(y)\frac{N_c^2-1}{(2\pi)^2}
\log^2\frac{\chi_2(y)}{\chi_1(y)}~.
\ee
Thus, at fixed impact parameter, the multiplicity scales as
$A_1^{1/3}$, up to the square of a logarithm of $(A_2/A_1)^{1/3}$.
For the transverse energy in the saturation regime one obtains
from~(\ref{sat_kt2})
\bea \label{Etsat}
\frac{\d E_\perp}{\d^2b_\perp\d y} &=&
g^4\sqrt{\frac{N_c}{2\pi}}\frac{N_c^2-1}{(2\pi)^2} \chi_1(y)\nonumber\\
& & \hspace{-1cm} \left(
\sqrt{\chi_1(y)}-\sqrt{\chi_2(y)}+\frac{\sqrt{\chi_2(y)}}{2}\log
\frac{\chi_2(y)}{\chi_1(y)} \right)~.
\eea
In the saturation regime~(\ref{dNdy_sat},\ref{Etsat})
as well as in the perturbative regime~(\ref{dNdy_pert},\ref{Etpert}) the
transverse energy per gluon is practically independent of $A_1$, while
a weak increase $\propto A_2^{1/6}$ is expected.

The average transverse momentum in the saturation regime follows
from~(\ref{dNdy_sat},\ref{Etsat}):
\be
\langle k_\perp\rangle = 2\; Q_s^{(2)}\;
\frac{\xi-1-\log\xi}{\log^2\xi}~,   \label{mkperp_sat1}
\ee
where $\xi(y)=\sqrt{\chi_1(y)/\chi_2(y)}
=Q_s^{(1)}(y)/Q_s^{(2)}(y)$. 
From dimensional considerations, it has been suggested~\cite{juergen} that in
symmetric $A+A$ collisions, and at
central rapidity, $\langle k_\perp\rangle^2$ scales with the multiplicity per
unit of transverse area and of rapidity,
\be
\langle k_\perp\rangle^2 \propto \frac{\d N}{\d^2b_\perp\d y}~.
\ee
A similar scaling relation can be derived from
eqs.~(\ref{dNdy_sat},\ref{mkperp_sat1}) for the asymmetric case,
\be \label{mkperp}
\langle k_\perp\rangle^2 \propto 
\frac{\d N}{\d^2b_\perp\d y} \; \frac{g^2}{\xi^2} \;
\frac{\left(\xi-1-\log\xi\right)^2}{\log^6\xi}~.
\ee
Thus, $\langle k_\perp\rangle^2$ is
proportional to the multiplicity per unit of rapidity and
transverse area, times a function of the ratio of the saturation momenta.
If source one is very much weaker than source two, i.e.\ in the limit
$|\log\xi|\gg1-\xi$, the third factor on the
right-hand-side of~(\ref{mkperp}) depends on $\log\xi$ only.
Neglecting that dependence, and assuming as before that $\chi_{1,2}$ are
proportional to $A_{1,2}^{1/3}$, one has the approximate scaling relation
\be \label{simple_scaling}
\langle k_\perp\rangle^2 \propto \left(\frac{A_2}{A_1}\right)^{1/3}
\frac{\d N}{\d^2b_\perp\d y}\propto
\left(\frac{1}{A_1 A_2}\right)^{1/3} \frac{\d N}{\d y}~.
\ee
In practice though we expect significant corrections to the simple scaling
relation~(\ref{simple_scaling}), as given by eq.~(\ref{mkperp}).

\begin{figure}[htp]
\centerline{\hbox{\epsfig{figure=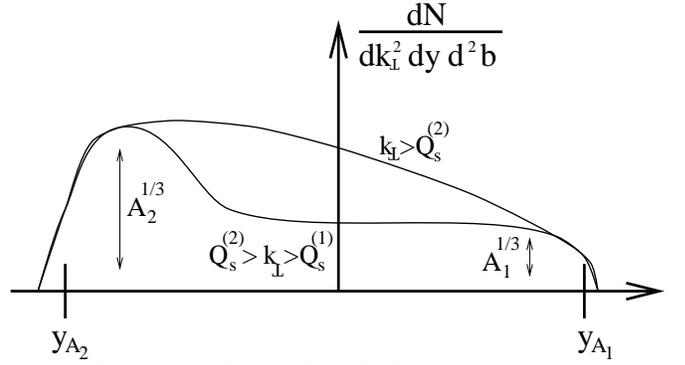,width=8.8cm}}}
\caption{Schematic rapidity distribution for particles produced in
high-energy $p+A$ collisions (or, more generally, for particles produced
in $A_1+A_2$ collisions at $A_1\ll A_2$).
The upper curve refers to the perturbative regime, the lower curve refers to
$k_\perp$ between the saturation scales for the two sources.}
\label{figy}
\end{figure}

Qualitatively, the rapidity distribution predicted from eq.~(\ref{sat_kt2})
is as follows\footnote{A quantitative computation requires to solve for the
RG evolution of the $\chi$'s first, which is out of the scope of the present
manuscript.}, see Fig.~\ref{figy}.
For rapidities far from the fragmentation region of the large
nucleus, and for
$g^4N_c\chi_1(y)/8\pi \lton k_\perp^2\lton g^4N_c\chi_2(y)/8\pi$,
$\d N/\d^2k_\perp\d y$ varies like
\be
\frac{\d^2 N}{\d y^2} \propto g^2 \chi_1'(y)~, 
\ee
where we have suppressed the dependence on transverse momentum, which is
supposed to be held fixed somewhere within the saturation regime.
Thus, an experimental measure for the RG evolution of the CGC density
parameter is
\be \label{RGchi1}
\frac{\d\log \d N/\d y}{\d y} =
\frac{\d\log\chi_1(y)}{\d y}~.
\ee

Now consider the case of high $k_\perp^2 > g^4N_c\chi_2(y)/8\pi$ described
by eq.~(\ref{pert_kt4}). In that regime the rapidity distribution is
proportional to $\chi_1(y)\chi_2(y)$, and so $\d N/\d y\d^2k_\perp$ varies
with rapidity like
\be \label{RGchi12}
\frac{\d\log \d N/\d y}{\d y} = \frac{\d\log\chi_1(y)}{\d y}
  + \frac{\d\log\chi_2(y)}{\d y}~.
\ee
Subtracting~(\ref{RGchi1}) from~(\ref{RGchi12}), that is
$(\d^2 N/\d y^2)/(\d N/\d y)$ measured at small transverse momentum
from that at larger transverse momentum, provides an experimental
measure for the RG evolution of $\chi_2(y)$.

\section{Summary} \label{summary}
In summary, we have computed the radiation field produced in the collision
of two ultrarelativistic, non-abelian, classical color charge sources
for the case where one of the sources is much stronger than the other.
Accordingly, we have linearized the Yang-Mills equations in
field 1, but solved them to all orders in field 2. The renormalization-group
evolution of the color charge density $\chi$ is not dropped.
This problem is relevant for $p+A$ collisions at high energy, or more
generally for $A_1+A_2$ nuclear collisions where $A_1\ll A_2$, or even for
symmetric $A+A$ collisions at large rapidities far away from $y=0$
(i.e.\ midrapidity) where the RG evolution ensures that the effective
$\chi_1$ is much smaller than $\chi_2$.

We obtain the following results relevant for phenomenology. At high transverse
momenta, $k_\perp^2 > g^4 N_c \chi_2/8\pi$, the distribution in $k_\perp$ is
proportional to the standard $\chi_1\chi_2/k_\perp^4$ known from perturbation
theory. Thus, the unintegrated distribution scales like $(A_1 A_2)^{1/3}$.
The total contribution from high transverse momenta, integrated over $k_\perp$
and impact parameters $b_\perp$, scales like $A_1^{1/3} A_2^{2/3}$.

In the saturation region 
$g^4 N_c \chi_2/8\pi > k_\perp^2 > g^4 N_c \chi_1/8\pi$, the distribution is
proportional to $\chi_1/k_\perp^2$; it decreases much less quickly with
transverse momentum than the result from perturbation theory. This may 
in principle provide
experimental information as to the value of $\chi_2(y)$.
At fixed $k_\perp^2$, the gluon distribution
scales like $A_1^{1/3}$ for fixed impact parameter,
or like $A_1^{1/3} A_2^{2/3}$ when one integrates over $\d^2b_\perp$.
Note that, up to logarithmic corrections,
the $k_\perp$-integrated distribution scales {\em in exactly the
same way} with $A_2$ as in the perturbative regime
(no matter whether impact parameter selected or
integrated). In contrast, at fixed $k_\perp$ and $b_\perp$ the multiplicity
in the perturbative regime scales as $A_2^{1/3}$ while in the saturation
regime it is independent of $A_2$ (up to a logarithm)~!
(Or, without impact parameter selection, we have a scaling with
$A_2$ in the perturbative regime versus scaling with
$A_2^{2/3}$ in the saturation region.)

Furthermore, at fixed transverse momentum, the quantity
$\d\log (\d N/\d y)/\d y$ allows an {\em experimental} measurement
of the RG evolution of the color charge density parameter $\chi$, and a
check whether the saturation regime has been reached.
The slope of $\d N/\d y$ at rapidities far from the fragmentation
region of the large nucleus measures the RG evolution of $\chi_1$. Also,
subtracting the slope of the $\d N/\d y$
measured at small transverse momentum
(within the saturation regime)
from that at larger transverse momentum
(in the perturbative regime), provides experimental access to
the RG evolution of $\chi_2$, and for its $A$ dependence.
Such differential measurements at RHIC and LHC should provide insight regarding
high-density QCD and the properties of the CGC, for example the value of
its fundamental parameter $\chi$ and its RG evolution (in rapidity).

\acknowledgements
A.D.\ acknowledges helpful discussions with B.\ Jacak,
J.\ Jalilian-Marian, Y.\ Kovchegov, J.\ Schaffner,
R.\ Venugopalan, and also thanks
R.\ Venugopalan for a careful reading of the manuscript.
A.D.\ is grateful for support from the DOE Research Grant,
Contract No.\ DE-FG-02-93ER-40764.
This manuscript has been authored under contract No.\ DE-AC02-98CH10886
with the U.S.\ Department of Energy.


\begin{references}
\bibitem{McLerran:1994ni}
L.~McLerran and R.~Venugopalan, Phys.\ Rev.\ {\bf D 49}, 2233 (1994);
%[hep-ph/9309289].
%%CITATION = HEP-PH 9309289;%%
Phys.\ Rev.\ {\bf D 49}, 3352 (1994).
%[hep-ph/9311205].
%%CITATION = HEP-PH 9311205;%%

\bibitem{KV}
A.~Krasnitz and R.~Venugopalan, Nucl.\ Phys.\ B {\bf 557}, 237 (1999);
%[hep-ph/9809433].
%%CITATION = HEP-PH 9809433;%%
Phys.\ Rev.\ Lett.\  {\bf 84}, 4309 (2000);
%[hep-ph/9909203].
%%CITATION = HEP-PH 9909203;%%
Phys.\ Rev.\ Lett.\  {\bf 86}, 1717 (2001).
%[hep-ph/0007108].
%%CITATION = HEP-PH 0007108;%% 

\bibitem{Kovner:1995ts}
A.~Kovner, L.~McLerran and H.~Weigert, Phys.\ Rev.\ {\bf D 52}, 3809 (1995);
%[hep-ph/9505320].
%%CITATION = HEP-PH 9505320;%%
Phys.\ Rev.\ {\bf D 52}, 6231 (1995).
%[hep-ph/9502289].
%%CITATION = HEP-PH 9502289;%%

\bibitem{Gyulassy:1997vt}
M.~Gyulassy and L.~McLerran,
%``Yang-Mills radiation in ultrarelativistic nuclear collisions,''
Phys.\ Rev.\ C {\bf 56}, 2219 (1997).
%[nucl-th/9704034].
%%CITATION = NUCL-TH 9704034;%%

\bibitem{Kovchegov:1997ke}
Y.~V.~Kovchegov and D.~H.~Rischke,
Phys.\ Rev.\ C {\bf 56}, 1084 (1997).
%[hep-ph/9704201].
%%CITATION = HEP-PH 9704201;%%

\bibitem{Mueller:1999wm}
A.~H.~Mueller,
%``Parton saturation at small x and in large nuclei,''
Nucl.\ Phys.\ B {\bf 558}, 285 (1999).
%[hep-ph/9904404].
%%CITATION = HEP-PH 9904404;%%

\bibitem{Kovchegov:1998bi}
Y.~V.~Kovchegov and A.~H.~Mueller,
Nucl.\ Phys.\ B {\bf 529}, 451 (1998);
%[hep-ph/9802440].
%%CITATION = HEP-PH 9802440;%%
Y.~V.~Kovchegov, hep-ph/0011252.
%%CITATION = HEP-PH 0011252;%%

\bibitem{Iancu:2001ad}
J.~Jalilian-Marian, A.~Kovner, A.~Leonidov and H.~Weigert,
Phys.\ Rev.\ D {\bf 59}, 014014 (1999);
%[hep-ph/9706377].
%%CITATION = HEP-PH 9706377;%%
J.~Jalilian-Marian, A.~Kovner and H.~Weigert,
Phys.\ Rev.\ D {\bf 59}, 014015 (1999);
%[hep-ph/9709432].
%%CITATION = HEP-PH 9709432;%%
E.~Iancu, A.~Leonidov and L.~McLerran,
%``Nonlinear gluon evolution in the color glass condensate. I,''
hep-ph/0011241;
%%CITATION = HEP-PH 0011241;%%
%E.~Iancu, A.~Leonidov and L.~McLerran,
%``The renormalization group equation for the color glass condensate,''
Phys.\ Lett.\ B {\bf 510}, 133 (2001);
%%CITATION = HEP-PH 0102009;%%
E.~Iancu and L.~McLerran,
%``Saturation and universality in QCD at small x,''
Phys.\ Lett.\ B {\bf 510}, 145 (2001).
%%CITATION = HEP-PH 0103032;%%

\bibitem{Jalilian-Marian:1997xn}
J.~Jalilian-Marian, A.~Kovner, L.~McLerran and H.~Weigert,
Phys.\ Rev.\ {\bf D 55}, 5414 (1997).
%[hep-ph/9606337].
%%CITATION = HEP-PH 9606337;%%

\bibitem{Guo:1999pe}
X.~Guo,
Phys.\ Rev.\ D {\bf 59}, 094017 (1999).
%[hep-ph/9812257].
%%CITATION = HEP-PH 9812257;%%

\bibitem{Balitsky:1996ub}
I.\ Balitsky,
%``Operator expansion for high-energy scattering,''
Nucl.\ Phys.\ B {\bf 463}, 99 (1996);
%[hep-ph/9509348].
%%CITATION = HEP-PH 9509348;%%
Y.~V.\ Kov\-che\-gov,
Phys.\ Rev.\ D {\bf 60}, 034008 (1999).
%[hep-ph/9901281].
%%CITATION = HEP-PH 9901281;%%

\bibitem{blaizot}
J.~P.~Blaizot and A.~H.~Mueller,
%``The Early Stage Of Ultrarelativistic Heavy Ion Collisions,''
Nucl.\ Phys.\ B {\bf 289}, 847 (1987);
%%CITATION = NUPHA,B289,847;%%
K.~J.~Eskola, K.~Kajantie, P.~V.~Ruuskanen and K.~Tuominen,
Nucl.\ Phys.\ B {\bf 570}, 379 (2000);
%[hep-ph/9909456].
%%CITATION = HEP-PH 9909456;%%
X.~Wang and M.~Gyulassy,
%``Energy and centrality dependence of rapidity densities at RHIC,''
Phys.\ Rev.\ Lett.\  {\bf 86}, 3496 (2001);
%[nucl-th/0008014].
%%CITATION = NUCL-TH 0008014;%%
D.~Kharzeev and M.~Nardi,
%``Hadron production in nuclear collisions at RHIC and high density QCD,''
Phys.\ Lett.\ B {\bf 507}, 121 (2001);
%[nucl-th/0012025].
%%CITATION = NUCL-TH 0012025;%%
H.~J.~Drescher, M.~Hladik, S.~Ostapchenko, T.~Pierog and K.~Werner,
%``Parton-based Gribov-Regge theory,''
hep-ph/0007198;
%%CITATION = HEP-PH 0007198;%%
N.~Armesto and C.~A.~Salgado,
%``A geometrical estimation of saturation of partonic densities,''
hep-ph/0011352.
%%CITATION = HEP-PH 0011352;%%

\bibitem{juergen}
L.~McLerran and J.~Schaffner-Bielich,
Phys.\ Lett.\ B {\bf 514}, 29 (2001);
%%CITATION = HEP-PH 0101133;%%
J.~Schaffner-Bielich, D.~Kharzeev, L.~D.~McLerran and R.~Venugopalan,
nucl-th/0108048.
%%CITATION = NUCL-TH 0108048;%%
\end{references}
\end{document}